\def\inv{^{\raise.15ex\hbox{${
  \scriptscriptstyle -}$}\kern-.05em 1}}

\def\Dsl{\,\raise.15ex\hbox{$/$}\mkern-13.5mu D}
\def\dsl{\raise.15ex\hbox{$/$}\kern-.57em\hbox{$\partial$}}

\def\lspace{\ifx\answ\bigans{}\else\qquad\fi}

\def\CR{\hbox{{$\cal R$}}}

\def\CO{\hbox{{$\cal O$}}}

\def\lform{\hbox{$\sqcup$}\llap{\hbox{$\sqcap$}}}
\def\darr#1{\raise1.5ex\hbox{$\leftrightarrow$}
\mkern-16.5mu #1}

\def\h{{{1\over2}}}

\def\INT{{\textstyle \int\kern-.642em\int}}
\def\R{{R\kern-.647em R}}
\def\C{{C\kern-.647em C}}

\def\T{{T\kern-.647em T}}
\def\Q{{Q\kern-.647em Q}}
\def\F{{F\kern-.647em F}}
\def\Z{{Z\kern-.647em Z}}
\def\N{{N\kern-.79em N}}

\def\eps{{\epsilon}}

\def\trace{{\rm Tr\, }}

\def\tens{\mathop{\otimes}}

\def\la{{\triangleright}}

\def\isom{{\cong}}

\def\Ad{{\rm Ad}}

\def\id{{\rm id}}

\def\nquad{{\!\!\!\!\!\!}}
\def\nqquad{\nquad\nquad}
\def\eqn#1#2{\begin{equation}#2\label{#1}\end{equation}}

\def\o{{}_{(1)}}\def\t{{}_{(2)}}

\def\und#1{{\underline {#1}}}

\def\uo{{{}^{(1)}}}\def\ut{{{}^{(2)}}}

\def\new#1{\goodbreak\goodbreak\bigskip
\noindent{\bf #1}}

\def\text#1{\mbox{\rm #1}}
\def\note#1{}

\def\frac#1#2{{{#1\over#2}}}

\def\proof{\goodbreak\noindent{\bf Proof\quad}}
\def\endproof{{\ $\lform$}\bigskip }

\def\align#1{\begin{eqnarray*}#1\end{eqnarray*}}

\def\und#1{{\underline{#1}}}

\def\vect{{\bf t}}\def\vecu{{\bf u}}
\def\vecx{{\bf x}}\def\vecv{{\bf v}}\def\vecb{{\bf b}}
\def\BDET{{\rm BDET}}

\documentstyle[twoside]{article}

\def\thebibliography#1{\section*{{\rm REFERENCES}}\list
 {[\arabic{enumi}]}{\settowidth\labelwidth{[#1]}\leftmargin\labelwidth
 \advance\leftmargin\labelsep
 \usecounter{enumi}}
 \def\newblock{\hskip .11em plus .33em minus -.07em}
 \sloppy
 \sfcode`\.=1000\relax}

\newtheorem{lemma}{Lemma}[section]
\newtheorem{propos}[lemma]{Proposition}
\newtheorem{example}[lemma]{Example}
\newtheorem{theorem}[lemma]{Theorem}

\begin{document}\baselineskip 21pt

{\ }\vspace{.8in}

\begin{center} {\bf QUANTUM AND BRAIDED LINEAR ALGEBRA}\footnote{1991
Mathematics Subject Classification 18D35, 16W30, 57M25, 81R50, 17B37.\note{str
in cats, Hopf alg, knots,  qg in qft, qg in defn}
}


\baselineskip 13pt{\ }\\ {\ }
\\ Shahn Majid\footnote{SERC Research Fellow and Drapers Fellow of Pembroke
College, Cambridge}
\end{center}
\vspace{10pt}

\begin{quote}\baselineskip 13pt
\noindent{ABSTRACT} Quantum matrices $A(R)$ are known for every $R$ matrix
obeying the Quantum Yang-Baxter Equations. It is also known that these act on
`vectors' given by the corresponding Zamalodchikov algebra. We develop this
interpretation in detail, distinguishing between two forms of this algebra,
$V(R)$ (vectors) and $V^*(R)$ (covectors). $A(R)\to V(R_{21})\tens V^*(R)$ is
an algebra homomorphism (i.e. quantum matrices are realized by the tensor
product of a quantum vector with a quantum covector), while the inner product
of a quantum covector with a quantum vector
transforms as a scaler. We show that if $V(R)$ and $V^*(R)$ are endowed with
the necessary braid statistics $\Psi$ then their braided tensor-product
$V(R)\und\tens V^*(R)$ is a realization of the braided matrices $B(R)$
introduced previously, while their inner product leads to an invariant quantum
trace. Introducing braid statistics in this way leads to a fully covariant
quantum (braided) linear algebra. The braided groups obtained from $B(R)$ act
on themselves by conjugation in a way impossible for the quantum groups
obtained from $A(R)$.
\end{quote}

\begin{quote}\baselineskip 13pt
\noindent{RESUM\'E}
Les matrices quantiques $A(R)$ sont connus pour chaque matrice $R$ qui
satisifie les equations de Yang-Baxter. Il est encore connu qu`ils agissent
sur les `vecteurs' donn\'es par l'alg\`ebre de Zamalodchikov correspondant.
Nous prolongons cette interpretation, distinguissant deux versions de cette
alg\'ebre, $V(R)$ (vecteurs) at $V^*(R)$ (covecteurs). $A(R)\to V(R_{21})\tens
V^*(R)$ est une homomorphisme des alg\`ebres, et le produit int\'erieur d'un
covecteur quantique avec un vecteur
quantique se transforme comme un scaleur. Nous demonstrons que si $V(R)$ et
$V^*(R)$ sont munis des statistiques tress\'ees $\Psi$, alors leur produit
tensoriel-tress\'e  $V(R)\und\tens V^*(R)$ est une r\'ealization des matrices
tress\'es $B(R)$ introduits d\'eja, et leur produit int\'erieur s'amene \`a une
trace invariante. Par introduisant les statistiques tress\'ees dans cette fa\c
con nous obtenons un alg\`ebre lin\'eair quantique (tress\'e) et totalement
covariant. Les groupes tress\'es obtenus de $B(R)$ s'agissent sur eux-m\^eme
par conjugaison dans une mani\`ere qui est impossible pour les groups
quantiques obtenus de $A(R)$.
\end{quote}

\newpage
\baselineskip 21pt

\section{Introduction}

Quantum matrices and groups have arisen in physics and it is well established
that they play an important role in certain physical theories. They also
suggest a new kind of quantum calculus (within the context of non-commutative
geometry) describing such physics. One aspect of the physics which is not,
however, so well covered by quantum groups is the braid-statistics of the
quantum fields. Here the non-commutativity arises not due to quantum effects
but due to non-trivial statistics (such as fermionic or anyonic statistics) and
suggests a kind of braided calculus as well as a quantum one.

In this paper we study the quantum and braided linear algebra associated to a
regular matrix $R\in M_n\tens M_n$ obeying the Quantum Yang-Baxter Equations
(QYBE). We begin with the quantum case, where quantum matrices of type $R$ are
defined as the bialgebra $A(R)$ of \cite{FRT:lie}, and clarify the precise way
that this `acts' on quantum vectors and quantum covectors. These are given by
variants of the Zamalodchikov algebra associated to $R$ as explained in
\cite[Sec. 6.3.2]{Ma:qua}, but more care is needed now to distinguish their
transformation properties. It can be expected that these considerations of
quantum linear algebra will ultimately shed some light also on more complex
constructions in the matrix form of quantum differential calculus as in
\cite{Zum:int}. For example, the quantum traces needed there arise in a
particularly obvious way from our considerations.

After this warm-up with quantum linear algebra we proceed to `braided linear
algebra'. Here the role of braided matrices is played by $B(R)$ introduced by
the author in \cite{Ma:exa}. The only difference between the $A(R)$ and the
$B(R)$, i.e. between quantum and braided linear algebra is that the latter is
fully covariant under an underlying quantum group
(which induces on it a braiding). Thus braided linear algebra means nothing
other than covariant quantum linear algebra.
Let $H$ be a fixed quantum group (with universal $R$-matrix in the sense of
\cite{Dri}). Then an object is $H$-covariant if $H$ acts on it in a way that
preserves all its structure. For example, an algebra $V$ is $H$-covariant if
$H$ acts on $V$ (and hence on $V\tens V$ via the comultiplication
$\Delta(H)\subset H\tens H$) and the multiplication is an intertwiner
\eqn{inv}{V\tens V{\buildrel\cdot\over\to} V ,\qquad h\la(a\cdot
b)=\cdot(h\la(a\tens b)),\qquad h\la 1=\eps(h)1}
where $\la$ is the relevant action. One says that $V$ is an $H$-module algebra.
Not only algebras but all quantum group constructions can be done fully
$H$-covariantly (one says that the constructions take place in the braided
category of $H$-representations.) This is the setting behind \cite{Ma:bra}. For
example, a super-quantum group is nothing other than a $\Z_2'$-covariant
quantum group
(where $\Z_2'$ denotes the group algebra of the group with two elements,
equipped with a certain non-standard quantum group structure, and its action
just corresponds to the grading).

The relevance of this notion to the present paper is that behind the bialgebras
$A(R)$ there is a quantum group (with universal $R$-matrix). For the standard
$R$ matrices this is $U_q(g)$, but note that we will not be tied to the
standard case below. The quantum group acts on $A(R)$ by a quantum coadjoint
action\cite[Theorem 3.2]{Ma:seq}\cite[Sec. 6]{Ma:exa} but this action does {\em
not} leave the defining relations
\eqn{FRT}{R\vect_1\vect_2=\vect_2\vect_1R}
invariant, i.e. $A(R)$ is not covariant in the way explained, even as an
algebra unless it is commutative (here $\vect_1,\vect_2$ denote the matrix of
generators $\vect$ viewed in $M_n\tens M_n$ in the standard way). The idea
behind $B(R)$ is that the relations (\ref{FRT}) must be modified in a certain
way to restore covariance. There is a canonical way to do this, namely a
process of {\em transmutation} that turns a suitable quantum group into one
that is covariant (for example it can be used to superize or anyonize suitable
quantum groups)\cite{Ma:tra}. In our case the transmutation of $A(R)$ gives
$B(R)$ as generated by $n^2$ elements $u^i{}_j$ (and 1) with relations and
coalgebra\cite{Ma:exa}
\eqn{B(R)}{R_{21}\vecu_1 R_{12} \vecu_2=\vecu_2 R_{21} \vecu_1
R_{12},\qquad\quad \und\Delta u^i{}_j=u^i{}_k\tens u^k{}_j,\qquad \und\eps
u^i{}_j=\delta^i{}_j.}
The relations were written with all the four $R$'s on the right in
\cite{Ma:exa} but a close inspection of the indices will show that (\ref{B(R)})
is just the same. Note also that these relations (\ref{B(R)}) are known in
quite another context, namely for the ordinary quantum groups $U_q(g)$ in the
form with generators $L=l^+Sl^-$. The reason for this is in fact an accident of
the particular `self-dual' structure of $U_q(g)$ as we explain in detail in
\cite{Ma:skl}. In general $B(R)$ is quite different as an algebra from any
quantum group, especially in the triangular case when $R_{21}R_{12}=1$, and its
conceptual meaning is also quite different because its role is as a braided or
covariant version of the quantum groups of function algebra type, not at all of
enveloping algebra type. Nevertheless for $U_q(g)$ we can certainly exploit
this accident to apply some of our results below about $B(R)$ to obtain
information about the covariance properties of $U_q(g)$ also.

These covariance properties of $B(R)$ were explained in detail in \cite{Ma:exa}
where we gave the coadjoint actions of the underlying quantum group etc. On the
other hand, this underlying quantum group can be hard to compute in practice
when $R$ is not a standard one. This can be avoided if we speak of (\ref{inv})
not in terms of a quantum group $H$ acting but in a dual form, in terms of the
{\em coaction} of a dual quantum group $A$. A coaction is just like an action
but
 with arrows reversed. Thus it means a map $V\to V\tens A$ instead of $H\tens
V\to V$ (left actions correspond to right coactions of the dual). For the case
of $B(R)$ above the underlying dual quantum group with respect to which
everything is covariant, is nothing other than $A(R)$ itself (modulo
`determinant-type' relations to provide an honest antipode). Moreover,
coactions might seem a little unfamiliar but when it comes to coactions of
matrix dual quantum groups such as $A(R)$, they take a very simple `matrix'
form. This point of view has been stressed by Manin in \cite{Man:non} and has
also become popular in physics, e.g.\cite{Zum:int}. For this reason one of our
first goals, in Section~2, will be to carefully convert the $H$-covariance
conditions (\ref{inv}) as used in \cite{Ma:exa}\cite{Ma:bra}, into this
`matrix' form.

A careful study of this will lead also to our notion of quantum vectors $V(R)$
and covectors $V^*(R)$ of $R$ type, based on the Zamalodchikov algebra and both
fully covariant. We show how they can be used to realize the algebra $A(R)$
itself in the same way that a matrix can be decomposed into the rank one
matrices $\vert i><j\vert$, while $\sum <i\vert i>$ transforms as a scaler.
This has some similarities with Manin's realization\cite{Man:non}, see also
\cite{Sud:mat}, but represents in fact a different and `orthogonal' formalism
to that. This is evident from the simplest
example whre $R$ is the $SL_q(2)$ $R$-matrix, for then we find $V(R)=\C^{2\vert
0}_{q}$ and $V^*(R)=\C^{2\vert 0}_{q^{-1}}$ in Manin's notation: they are both
bosonic quantum planes. For another choice of normalization one can have both
fermionic, but we do not mix bosonic and fermionic quantum planes as in Manin's
approach. This is even more evident when $R$ has several (not only two)
relevant normalizations, which in our approach are not mixed (there is a
complete quantum linear algebra with both quantum vectors and quantum covectors
for each choice of normalization).

Finally, we will be ready in Section~3 to give braided (i.e. covariant)
versions of all these considerations with $B(R)$ recovered when $V(R)$ and
$V^*(R)$ are no longer mutually commutative but treated instead with braid
statistics $\Psi$. $B(R)$ acts on them, as well as on itself by conjugation. As
a spin-off we will recover from the above scaler a general formula for the
quantum trace, useful in other contexts. Section 4 is devoted to computing some
of the simplest
examples of the theory, including a non-standard one related to the 8-vertex
model. Another example makes partial  contact with some formulae of recent
interest in physics\cite{IsaMal:def}. The paper concludes in Section~5 with
some details of the relationship between $A(R)$ and $B(R)$ (i.e. of
transmutation), interpreting it as a kind of partition function with prescribed
boundary conditions. In addition, an appendix provides an abstract
(diagrammatic) proof of one of our main theorems, based on the
braided-commutativity of braided groups.

\new{Acknowledgments} I thank D. Gurevich for posing one of the problems solved
in this paper. Our Proposition~3.5 in Section 3 can be viewed as a braided
version of his result in the symmetric (but not braided) case in
\cite{Gur:alg}.

\section{Transformation of Vectors and Covectors}

In this preliminary section we begin by establishing the matrix description of
the transformation properties that we will need. Since there seems to be a gap
between the standard mathematical way of describing adjoint coactions etc and
the matrix notations preferred by physicists, we will  explain their
equivalence carefully (with proofs). Most probably this is well-known to
experts, but I didn't find an adequate treatment elsewhere.

Firstly, recall that a bialgebra means an algebra $A$ over a field or
commutative ring $k$ and a map $\Delta:A\to A\tens A$ which is an algebra
homomorphism and coassociative. There also needs to be a counit $\eps:A\to k$.
The matrix notation stems from the following well-known and innocent
observation. Let $A$ be an algebra with $n^2$ matrix generators
$\vect=(t^i{}_j)$. Suppose that $\eps(t^i{}_j)=\delta^i{}_j$ extends
multiplicatively to a map $\eps:A\to k$. Let $\Delta t^i{}_j=t^i{}_k\tens
t^k{}_j$ and $\Delta(1)=1\tens 1$. Then $(A,\eps,\Delta)$ is a bialgebra if and
only if the following holds: If $\vect,\vect'$ are two identical sets of
generators of $A$, mutually commuting elementwise, then $\vect''=\vect\vect'$
is also a realization of $A$ (i.e. $t''^i{}_j=t^i{}_k t'^k{}_j$ obey its
relations also). In truth, this observation  is just saying that the $\Delta$
that we desire is an algebra homomorphism
to $A\tens A$ (i.e. $A\tens A$ is a realization of $A$), where $A\tens A$ as an
algebra is of course given by two mutually commuting copies of $A$, i.e.
generated by $\vect=\vect\tens 1$ and $\vect'=1\tens \vect$ in $A\tens A$. The
main content of the notation is to omit writing the tensor product,
distinguishing the elements of the second factor instead by the prime.

We have gone through the rationale in detail because the same method of
omitting tensor products works for also comodule algebras. Thus, if $A$ is a
bialgebra and $\beta:V\to V\tens A$ a comodule (the dual notion of an action)
and $V$ is an algebra then $V\tens V\to V$ is $A$-covariant (in the sense of
(\ref{inv}) but in our dual language) if $\beta$ is an algebra homomorphism.
One says that $V$ is an $A$-comodule algebra. The reader can easily see that
this is just the condition in (\ref{inv}) with arrows reversed and left-right
interchanged. The following observation was probably first stressed by Manin in
\cite{Man:non} in connection with the quantum plane. It is surely also well
known to others.

\begin{lemma} Let $A$ be a matrix bialgebra (as above) and $V$ an algebra with
$n$ generators $\vecx=(x_i)$ (written as a row vector) and 1. Define
$\beta(1)=1\tens 1$ and $\beta(x_j)=x_i\tens t^i{}_j$. Then $\beta$ makes $V$ a
comodule algebra if and only if the following holds: whenever $\vect$ is a copy
of the generators of $A$, $\vecx$ a copy of those of $V$, commuting elementwise
with the $\vect$, then $\vecx'=\vecx\vect$ is a realization of $V$.
\end{lemma}
\proof Here $t^i{}_j=1\tens t^i{}_j$ and $x_i=x_i\tens 1$ are the generators of
the tensor product algebra $V\tens A$
built from mutually commuting copies of $A$ and $V$. The condition is just that
the products $x'_j=x_it^i{}_j$ are a realization of $V$, i.e. that $\beta:V\to
V\tens A$ as defined is an algebra map. On the other hand, $\beta$ as defined
is already a
right coaction from the form of its definition. This is because to be a
coaction one needs $(\beta\tens\id)\beta=(\id\tens\Delta)\beta$ and
$(\id\tens\eps)\beta=\id$, which we see automatically as $\beta(x_i)\tens
t^i{}_j=x_{i'}\tens t^{i'}{}_i\tens t^i{}_j=x_{i'}\tens\Delta(t^{i'}{}_j)$ and
$x_i\eps(t^i{}_j)=x_j$ due to the matrix form of $\Delta,\eps$.
\endproof

So far we have only said carefully what is well-known. But the same method also
gives

\begin{lemma} Let $A$ be a matrix Hopf algebra (as above but with an antipode
$S$) and $V$ an algebra with  $n$ generators $\vecv=(v^i)$ (written as a column
vector) and 1. Define
$\beta(1)=1\tens 1$ and $\beta(v^j)=v^i\tens S t^j{}_i$. Then $\beta$ makes $V$
a comodule algebra if and only if the following holds: whenever $\vect$ is a
copy of the generators of $A$, $\vecv$ a copy of those of $V$, commuting
elementwise with the $\vect$, then $\vecv'=\vect^{-1}\vecv$ is a realization of
$V$. Here $\vect^{-1}=S\vect$, i.e., the matrix with entries $(St^i{}_j)$.
\end{lemma}
\proof Here $t^i{}_j=1\tens t^i{}_j$ (as before) and $v^i= v^i\tens 1$ are the
generators of the tensor product algebra $V\tens A$ built from mutually
commuting copies of $A$ and $V$. We use the fact that they mutually commute in
the tensor product to write the $S\vect$ on the left even though it lives in
the second factor of $V\tens A$. The condition is just that the
$v'^i=St^i{}_jv^j$ is a realization of $V$, i.e. that $\beta:V\to V\tens A$ as
defined is an algebra map. Once again, $\beta$ as defined is already a right
coaction from the form of its definition. This is because $\beta(v^j)\tens
St^i{}_j=v^{j'}\tens St^{j}{}_{j'}\tens
St^i{}_j=v^{j'}\tens\Delta(St^{i}{}_{j'})$ and $v^i\eps(St^j{}_i)=v^j$ due to
the matrix form of $\Delta,\eps$ and that $S$ is an anti-coalgebra map while
$\eps\circ S=\eps$.
\endproof

We have, combining these,

\begin{lemma}  Let $A$ be a matrix Hopf algebra (as above) and $V$ an algebra
with  $n^2$ generators $\vecb=(b^i{}_j)$ and 1. Define
$\beta(1)=1\tens 1$ and $\beta(b^i{}_j)=b^m{}_n\tens (S t^i{}_m)t^n{}_j$. Then
$\beta$ makes $V$ a comodule algebra if and only if the following holds:
whenever $\vect$ is a copy of the generators of $A$, $\vecb$ a copy of those of
$V$, commuting with the $\vect$, then $\vecb'=\vect^{-1}\vecb\vect$ is a
realization of $V$.
\end{lemma}
\proof Here $t^i{}_j=1\tens t^i{}_j$ (as before) and $b^i{}_j=b^i{}_j\tens 1$
are the generators of the tensor product algebra $V\tens A$ built from mutually
commuting copies of $A$ and $V$. We again use the fact that they mutually
commute in the tensor product to write the $St^i{}_m$ part on the left even
though it lives in the second factor of $V\tens A$ along with the $t^n{}_j$
part. The condition is just that $b'^i{}_j=(St^i{}_m) b^m{}_n t^n{}_j$ is a
realization of $V$, i.e. that $\beta:V\to V\tens A$ as defined is an algebra
map. The map $\beta$ as defined is already a right coaction because
$\beta(b^m{}_n)\tens (St^i{}_m)t^n{}_j=b^{m'}{}_{n'}\tens
(St^{m}{}_{m'})t^{n'}{}_n\tens
(St^i{}_m)t^n{}_j=b^{m'}{}_{n'}\tens\Delta((St^{i}{}_{m'})t^{n'}{}_j)$ and
$b^m{}_n\eps((St^i{}_m)t^n{}_j)=b^i{}_j$ due to $\Delta,\eps$ being algebra
homomorphisms and the arguments already given in the proofs of the two lemmas
above.
\endproof

We now give some important (but not the only) examples of comodule algebras of
such type. Let $R\in M_n\tens M_n$ be a matrix solution of the QYBE and
$A=A(R)$ the FRT bialgebra (which is of the matrix type above). The following
two examples are variants of the Zamalodchikov algebra on $n$ generators and
the known coaction of $A(R)$ on it as shown for general $R$ in \cite[Sec.
6.3.2]{Ma:qua}. The new part lies in our careful and matching selection of
conventions for our present purposes. We fix a single invertible constant
$\lambda$ throughout (and a fixed normalization of $R$ which we do not change
further).

\begin{example} We define $V^*(R)$ to be the algebra with $n$ generators $x_i$
and 1, and relations $x_ix_k= x_n x_m \lambda R^m{}_i{}^n{}_k $. Writing
$\vecx=(x_i)$ as a row vector and $\vect
=(t^i{}_j)$ as a matrix (with values in their respective algebras), the
assignment $\vecx'=\vecx\vect$
makes $V^*(R)$ into a right $A(R)$-comodule algebra. We call it the algebra of
quantum covectors of $R$-type.
\end{example}
\proof A proof in conventional comodule notation is in \cite[Sec.
6.3.2]{Ma:qua}. In our matrix notation it is simply as follows. The relations
of $V^*(R)$ are $\vecx_1\vecx_2=\vecx_2\vecx_1\lambda R$ where
$\vecx_1=\vecx\tens 1$ and $\vecx_2=1\tens\vecx$. We check
$\vecx'_1\vecx'_2=\vecx_1\vect_1\vecx_2\vect_2=\vecx_2\vecx_1 \lambda
R\vect_1\vect_2=\vecx_2\vecx_1 \lambda\vect_2\vect_1 R=\vecx'_2\vecx'_1\lambda
R$ so the transformed covectors obey the same relations. We used that the
$\vecx,\vect$ commute, the relations in $V^*(R)$ and the relations of $A(R)$.
\endproof

\begin{example} We define $V(R)$ to be the algebra with $n$ generators $v^i$
and 1, and relations $v^iv^k=\lambda R^i{}_j{}^k{}_l v^lv^j$. Writing
$\vecv=(v^i)$ as a column vector and $\vect^{-1}=(St^i{}_j)$ for the matrix
inverse of $\vect$ (with values in the respective algebras), the assignment
$\vecv'=\vect^{-1}\vecv$
makes $V(R)$ into a right $A$-comodule algebra, where $A$ is a suitable dual
quantum group obtained from $A(R)$. We call $V(R)$ the algebra of quantum
vectors of $R$-type.
\end{example}
\proof The proof is similar to the preceding example. In the matrix notation it
is as follows. The relations of $V(R)$ are $\vecv_1\vecv_2=\lambda
R\vecv_2\vecv_1$. Then
$\vecv'_1\vecv'_2=\vect_1^{-1}\vecv_1\vect_2^{-1}\vecv_2
=\vect_1^{-1}\vect_2^{-1}\lambda R\vecv_2\vecv_1$ $=
\lambda R\vect^{-1}_2\vect_1^{-1}\vecv_2\vecv_1=\lambda R\vecv_2'\vecv_1'$.
We used the relations for $A(R)$ in a form obtained by applying the
antipode $S$ to the relations (\ref{FRT}). We assume that $A$ can be
obtained from $A(R)$ (by imposing determinant-type relations or by
inverting a determinant etc) in a way consistent with the coaction.
This is true, for example, for $R$ matrices that are regular in the
sense of Section~3 below. \endproof

The (standard) rationale behind these examples is from non-commutative
geometry, as explained in detail in \cite[Sec. 6.3.2]{Ma:qua} specifically for
examples of the above type. The point is that entries of $\vect$ are
non-commutative versions of the tautological functions $t^i{}_j(A)=A^i{}_j$ for
actual matrices $A$, while similarly $v^i$ and $x_i$ are non-commutative
versions of $v^i({V})=V^i$ and $x_i(X)=X_i$ for actual column and row vectors
${V,X}$.
Thus $A(R),V(R),V^*(R)$ are non-commutative versions of
$C(M_n),C(\R^n),C(\R^n)$. The usual action $\R^n\times M_n\to\R^n$ for example
appears in terms of these as a right comodule algebra structure $C(\R^n)\to
C(\R^n)\tens C(M_n)$. It is just this structure which we keep in the quantum
setting (of type $R$). Note that this could for example happen as a result of
actual quantization of a commutative algebra of observables of systems on
$\R^n$ and $M_n$. In this case the $t^i{}_j$ etc are quantum observables and
become operators. Thus we can think of the $\vect,\vecx,\vecv$ as
operator-valued matrices, covectors, vectors, in spite of their origin as
quantized tautological functions
(this is a standard point of view in quantum mechanics).

This all seems very reasonable, but let us note that

\begin{lemma} The subalgebra of $V(R)\tens V^*(R)$ with generators 1 and
\[\vecv\vecx=\pmatrix{v^1x_1&\cdots &v^1x_n\cr \vdots&&\vdots\cr v^nx_1&\cdots
&v^nx_n}\]
(i.e. with a matrix of generators $v^ix_j$) is a right $A$-comodule under the
assignment $(\vecv\vecx)'=\vect^{-1}\vecv\vecx\vect$ but {\em not} in general a
right $A$-comodule algebra (they do not obey the right relations). Likewise,
$A(R)$ itself is a right $A$-comodule under the assignment
$\vect''=\vect'^{-1}\vect\vect'$
where $\vect'$ denotes the copy of the generators lying in the coacting dual
quantum group $A$. But it is {\em not} in general a right $A$-comodule algebra.
\end{lemma}
\proof For the first part, since $A$ coacts on $V(R)$ and on $V^*(R)$
separately as above, it must have a tensor product coaction on their tensor
product algebra. The problem is that this does not in general give a comodule
algebra. The fundamental reason for this is that $\vecx,\vecv$ commute in the
tensor product algebra, but $\vecx'=\vecx\vect,\vecv'=\vect^{-1}\vecv$ do not
generally commute because the matrix entries of $\vect,\vect^{-1}$ do not
generally commute. For the second part, let us note that every dual quantum
group $A$ (here $A(R)$ modulo determinant-type relations to make it a Hopf
algebra) coacts on
itself by the adjoint coaction. This is the dual notion to the action of any
quantum group on itself by the adjoint action. Just as the latter always
respects its own multiplication (in the sense of (\ref{inv})) so the adjoint
coaction always respects its own comultiplication (it is a comodule coalgebra).
This is true also for $A(R)$ as a comodule coalgebra. However, again because
the matrix entries are generally non commuting, we do not have in general a
comodule algebra. \endproof

Thus the situation is not quite as we would hope. One has nevertheless

\begin{propos} The assignment $\vect=\vecv\vecx$ ($t^i{}_j=v^ix_j$) is a
realization of $A(R)$ in the algebra $V(R_{21})\tens V^*(R)$, i.e. gives an
algebra homomorphism $A(R)\to V(R_{21})\tens V^*(R)$. Here $R_{21}$ denotes $R$
transposed in the usual way.
\end{propos}
\proof The relations for $V(R_{21})$ are $\vecv_2\vecv_1=\lambda R
\vecv_1\vecv_2$. The $\vecv$ commute with the $\vecx$ in the tensor product
algebra, so we have
$R\vecv_1\vecx_1\vecv_2\vecx_2=R\vecv_1\vecv_2\vecx_1\vecx_2
=\lambda^{-1}\vecv_2\vecv_1\vecx_1\vecx_2
=\lambda^{-1}\vecv_2\vecv_1\vecx_2\vecx_1\lambda R
=\vecv_2\vecx_2\vecv_1\vecx_1 R$
so that the $\vecv\vecx$ realise (\ref{FRT}). \endproof

Finally, we note that the element $\vecx\vecv=\sum_ix_iv^i$ in $V^*(R)\tens
V(R)$ is clearly invariant under these transformations of $\vecx,\vecv$ by the
usual computation (even though the entries may be non-commuting). In other
words, under the tensor product coaction on $V^*(R)\tens V(R)$, this element
maps to $\vecx\vecv\tens 1$ in $V^*(R)\tens V(R)\tens A$ (it is a fixed point).
Again this is what we would like, although let us remark that this $\vecx\vecv$
need not be central in the algebra $V^*(R)\tens V(R)$ (nor in $V^*(R)\tens
V(R_{21})$) which is a little worrying.

In summary we see that (with a little care because the matrix entries are
non-commuting), the set-up above has some of the usual features of linear
algebra. We say that an algebra $V$ transforms as a quantum covector if it is
an example of Lemma~2.1, transforms as a quantum vector if an example of
Lemma~2.2 and transforms as a quantum matrix if an example of Lemma~2.3. On the
other hand, there are a couple of alarming features,  where the most naive
expectations do not hold. In particular, $A(R)$ itself as well as the `quantum
rank-one matrices'
$\vecv\vecx$ do not exactly fulfill the conditions of Lemma~2.3, while the
obvious scaler element is not central.

The problems here are all attributable to the fact that $A(R)$, while it serves
well (in a quotient) as the quantum symmetry of the system, is not covariant
under itself. It seems that in the quantum universe, the role of
quantum symmetry (played by the dual quantum group) and quantum matrix (in the
sense of non-commutative geometry) living in that universe, become
disassociated. As explained in the introduction, the braided matrices $B(R)$
have been introduced in \cite{Ma:exa} precisely as a covariantized version of
$A(R)$ and better serve the latter role. We see this now in the next section.
For classical groups these two objects coincide.

\section{Braided Linear Algebra}

In this section we develop quantum linear algebra in a way that is fully
covariant under the dual quantum group $A$ given by $A(R)$ modulo
determinant-type relations. As explained in the last section, this plays the
hidden role of a symmetry but the role of matrices itself are played by $B(R)$.
We have

\begin{example} The algebra $B(R)$ with generators $\vecu=(u^i{}_j)$ and
relations in (\ref{B(R)}) forms an $A$-comodule algebra under the assignment
$\vecu'=\vect^{-1}\vecu\vect$ as in Lemma~2.3.
\end{example}
\proof This is the raison d'\^etre of the theory of braided groups. The
coaction originates as the adjoint coaction in Lemma~2.6 on $A(R)$, but the
relations of the latter were converted in \cite{Ma:eul} by a process of
transmutation to derive those of $B(R)$ as explained in \cite{Ma:exa}. For a
direct proof we have easily $R_{21}\vect_1^{-1}\vecu_1\vect_1
R_{12}\vect_2^{-1}\vecu_2\vect_2=R_{21}\vect_1^{-1}\vecu_1\vect_2^{-1}
R_{12}\vect_1\vecu_2\vect_2=\vect_2^{-1}\vect_1^{-1}R_{21}\vecu_1R_{12}
\vecu_2\vect_1\vect_2$. Here we used (\ref{FRT}) in various forms and freely
commuted $\vecu_1$ with $\vect_2$ etc (they live in different algebras and
in different matrix spaces). Applying (\ref{B(R)}) to the result we have
similarly $\vect_2^{-1}\vect_1^{-1}\vecu_2R_{21}\vecu_1
R_{12}\vect_1\vect_2=\vect_2^{-1}\vecu_2\vect_2R_{21}\vect^{-1}_1\vecu_1
\vect_1R_{12}$ so that the transformed $\vecu$ obey the same relations
(\ref{B(R)}). \endproof

This $B(R)$, however, is not a bialgebra in an ordinary sense. With the matrix
comultiplication in (\ref{B(R)}) it becomes one with braid statistics, i.e. we
call it the {\em braided matrices} of type $R$ (to distinguish it from $A(R)$).
It nevertheless transforms as a quantum matrix in the sense of Lemma~2.3. The
reason for the necessity of a braiding here is not an accident but a
fundamental feature of doing quantum linear algebra in a fully covariant way.
To explain this let us note that there are situations in linear algebra where
we have to make transpositions $V\tens W\to W\tens V$, yet when $V,W$ are
quantum vectors or covectors such as above, the ordinary transposition map is
{\em not} covariant. For example, under the usual transposition, $\vecx\tens
\vecv\mapsto\vecv\tens\vecx$
but after a transformation $\vecx'\tens\vecv'=\vecx\vect\tens\vect^{-1}\vecv\ne
\vect^{-1}\vecv\tens\vecx\vect=\vecv'\tens\vecx'$ precisely because the matrix
entries of $\vect,\vect^{-1}$ need not commute when they are multiplied
together (according to the definition of the tensor product coaction). In
covariant quantum linear algebra we are forced to introduce a non-standard
`transposition' $\Psi_{V,W}:V\tens W\to W\tens V$ which is covariant and still
obeys the rules
\eqn{psi}{\Psi_{V,W\tens Z}=\Psi_{V,Z}\Psi_{V,W},\qquad \Psi_{V\tens
W,Z}=\Psi_{V,Z}\Psi_{W,Z}}
for any three covariant objects. This means that it does truly behave like
transposition.
Moreover, this $\Psi$ is required to be functorial, meaning that it must
commute in an obvious way with any other covariant linear operations between
objects. For example, if we multiply elements in one of our covariant algebras,
and then `transpose' the resulting element with an element  in another
covariant algebra, the result is the same as first `transposing' the factors
with the third element, and then multiplying.
See \cite{Ma:exa} for more discussion. The main difference with ordinary
transposition is that for general $R$ we are forced to drop
$\Psi_{W,V}\Psi_{V,W}=\id_{V,W}$. This means that mathematically all our
objects live in a braided (or quasitensor) category and $\Psi$ is called the
{\em braiding} or quasisymmetry.

The category in our case is the category of $A$-comodules and the braiding
$\Psi$ exists if $A$ is dual quasitriangular (roughly speaking, the dual of a
quantum group with universal $R$-matrix). It is easy to see that $A(R)$ is dual
quasitriangular as a bialgebra for any $R$ obeying the QYBE (the essential
computations for this were first given in \cite[Sec. 3.2.3]{Ma:qua}). It is
however, not automatic that this dual quasitriangular structure projects to the
quotients that may be needed to obtain an honest Hopf algebra $A$ (or even that
the latter really exists at all). We call $R$ {\em regular} if indeed a
quotient of $A(R)$ becomes a Hopf algebra $A$ and $R$ extends to a dual
quasitriangular structure
$\CR:A\tens A\to k$ with $\CR(\vect_1\tens\vect_2)=R$. We showed in \cite[Sec.
3.2.3]{Ma:qua} (in some form) that this is formally speaking always true, but
sometimes these formal expressions can fail. One needs $R$ and various matrices
built from $R$ to be invertible. Needless to say, all the standard $R$ matrices
are regular in this way, but we do not limit ourselves to the standard case,
requiring only that $R$ is regular. Note that in this set-up based on
\cite[Sec. 3]{Ma:qua}, the normalization of $R$ is determined.

The braiding between quantum matrices and themselves is the same as for the
example of $B(R)$ and was already given in \cite{Ma:exa}. Likewise  for vectors
with vectors. We summarise these and the other combinations as follows.

\begin{propos} Let $\vecx,\vecv,\vecu$ be any $A$-comodule algebras of
covector, vector and matrix type in the sense of Lemmas~2.1-2.3. Their mutual
braiding is given by
\[ \Psi(x_i\tens x_j)=x_n\tens x_m R^m{}_i{}^n{}_j,\quad \Psi(v^i\tens
v^j)=R^i{}_m{}^j{}_n v^n\tens v^m\]
\[ \Psi(x_i\tens v^j)=\tilde R^m{}_i{}^j{}_n v^n\tens x_m,\quad \Psi(v^i\tens
x_j)=x_n\tens v^m R^{-1}{}^i{}_m{}^n{}_j\]
\[ \Psi(u^i{}_j\tens x_k)=x_m\tens u^a{}_b R^{-1}{}^i{}_a{}^m{}_n
R^b{}_j{}^n{}_k,\quad \Psi(x_k\tens u^i{}_j)=u^a{}_b\tens x_m\tilde
R^n{}_k{}^i{}_a R^m{}_n{}^b{}_j\]
\[ \Psi(u^i{}_j\tens v^k)=v^m\tens u^a{}_b R^i{}_a{}^n{}_m\tilde
R^b{}_j{}^k{}_n,\quad \Psi(v^k\tens u^i{}_j)=u^a{}_b\tens v^m R^k{}_n{}^i{}_a
R^{-1}{}^n{}_m{}^b{}_j\]
\[\Psi(u^i{}_j\tens u^k{}_l)=u^p{}_q\tens u^m{}_n  R^{i}{}_a{}^d{}_{p}
R^{-1}{}^a{}_{m}{}^{q}{}_b
R^{n}{}_c{}^b{}_{l} {\tilde R}^c{}_{j}{}^{k}{}_d\]
where $\tilde R=((R^{t_2})^{-1})^{t_2}$ and $t_2$ denotes transposition in the
last two indices.
\end{propos}
\proof The braidings in the proposition are  special cases of the braiding in
the category of $A$-comodules. The braiding $\Psi_{V,W}$ in general is given by
applying the comodule maps to each of $V,W$, transposing the resulting $V,W$ in
the usual way and applying the dual quasitriangular structure $\CR:A\tens A\to
k$ to the two $A$ factors. Thus  $\Psi(x_i\tens x_j)=\CR(t^m{}_i\tens
t^n{}_j)x_n\tens x_m$ which evaluates to the matrix $R$. Likewise
$\Psi(x_i\tens v^j)=\CR(t^m{}_i\tens St^j{}_n)v^n\tens x_m$, which evaluates to
the matrix $\tilde R$. The others are similar. This is the method by which
$\Psi(u^i{}_j\tens u^k{}_l)$ was initially obtained and then verified directly
in \cite{Ma:exa}. Likewise, we can verify directly that all the above extend to
products of the generators and to tensor products according to the desired
properties of a braiding. \endproof

As in Section~2 we can adopt a more compact notation in which $\tens$ is
omitted, so that it looks like an algebra product. The rationale behind this is
precisely the formation of braided tensor product algebras.

\begin{lemma} If $V,W$ are two $A$-comodule algebras then $V\und\tens W$
defined with multiplication
\[ (v\tens w)(u\tens z)=v\Psi(w\tens u)z,\quad v,u\in V,\ w,z\in W\]
is also an $A$-comodule algebra. Writing $v=v\tens 1,w=1\tens w$ (so that
$v\tens w=vw$)  the braided tensor product algebra structure has the relations
of $V$, the relations of $W$ and the cross relations $wu:=\Psi(w\tens u)$. We
call these cross relations (expressing $\Psi$) the {\em statistics relations}
between $V,W$.
\end{lemma}
\proof This is an elementary first step in the theory of algebras and Hopf
algebras in braided categories as in \cite{Ma:exa}\cite{Ma:bg}. Since the
multiplications of $V,W$ are covariant and $\Psi$ is also covariant, the
multiplication of $V\und\tens W$ must also be covariant, i.e. it is an
$A$-comodule algebra. That this multiplication is associative follows from
functoriality of $\Psi$ and its braid relations (\ref{psi}). \endproof

Thus when we use such statistics, the algebra structure on $B(R)\und\tens B(R)$
is different from the usual one, including now the effects of the statistics
$\Psi$. Only with respect to this is the comultiplication $\und\Delta:B(R)\to
B(R)\und\tens B(R)$ an algebra homomorphism\cite{Ma:exa}. In this respect then,
$B(R)$ resembles more a super-quantum group than an ordinary one, with even
more complicated statistics than in the super case. For other examples we note,

\begin{lemma} The statistics relations between quantum spaces of covector,
vector and matrix type (with generators $\vecx,\vecv,\vecu$ respectively) are
\[ \vecx_1\vecx_2:=\vecx_2\vecx_1R,\quad \vecv_1\vecv_2:=R\vecv_2\vecv_1,\quad
\vecx_1R\vecv_2:=\vecv_2\vecx_1,\quad \vecv_1\vecx_2:=\vecx_2R^{-1}\vecv_1\]
\[ \vecu_1\vecx_2:=\vecx_2 R^{-1}\vecu_1 R,\quad \vecv_1\vecu_2:=R\vecu_2
R^{-1}\vecv_1,\quad R^{-1}\vecu_1 R\vecv_2:=\vecv_2\vecu_1,\quad
\vecx_1R\vecu_2R^{-1}:=\vecu_2\vecx_1 \]
\[ R^{-1}\vecu_1 R\vecu_2:=\vecu_2 R^{-1}\vecu_1 R\]
The use of $:=$ is to stress that the right hand side is the definition of the
left hand side in the tensor product algebra (and not vice-versa if
$\Psi^2\ne\id$).
\end{lemma}
\proof This is simply Proposition~3.2 written in a compact form with the symbol
$\Psi$ omitted on the left of each equation and tensor products omitted. The
${}_{1,2}$ induced refer to matrix indices. Also, we have to be careful not to
use the above relations in the wrong way. The reverse ones, given by
$\Psi^{-1}$ are generally different. Thus, $\vecv_1\vecu_2:=R\vecu_2
R^{-1}\vecv_1$ should not be confused with $\vecv_2\vecu_1=:R^{-1}\vecu_1
R\vecv_2$. Another way to distinguish them is to label the elements of the
second algebra with a $'$ as explained in \cite[Sec. 2]{Ma:exa}.\endproof

Let us note the formal similarity between these statistics relations and the
algebra defining relations in the examples of $V(R),V^*(R),B(R)$. This
similarity reflects the sense in which these algebras are all
braided-commutative \cite{Ma:eul}\cite{Ma:bg}\cite{Ma:exa}. We are now ready to
give braided analogs of the results of Section~2. From Lemma~3.3 we know that
$V(R)\und\tens V^*(R)$ is an $A$-comodule algebra (i.e. transforms as a quantum
matrix) -- so long as we use the braided tensor product algebra there is no
problem such as in Lemma~2.6.

\begin{propos} The assignment $\vecu=\vecv\vecx=\pmatrix{v^1x_1&\cdots
&v^1x_n\cr \vdots&&\vdots\cr v^nx_1&\cdots &v^nx_n}$
is a realization of $B(R)$ in $V(R)\und\tens V^*(R)$, i.e. gives  a (covariant)
algebra homomorphism $B(R)\to V(R)\und\tens V^*(R)$, where the latter is the
braided tensor product algebra.
\end{propos}
\proof  We compute $R_{21}\vecv_1\vecx_1
R_{12}\vecv_2\vecx_2:=R_{21}\vecv_1\vecv_2\vecx_1\vecx_2=\vecv_2\vecv_1
\lambda^{-1}\vecx_1\vecx_2=\vecv_2\vecv_1\vecx_2\vecx_1R_{12}
=:\vecv_2\vecx_2R_{21}\vecv_1\vecx_1R_{12}$.
The first and last equalities use the third statistics relation
displayed in the preceding lemma (of the form $\Psi(x_i\tens v^j)$).
The middle equalities use the defining relations in the algebras
$V(R),V^*(R)$. Hence $\vecv\vecx$ is a realization of the braided
matrices $B(R)$.  Moreover, this realization is manifestly covariant,
so that (by functoriality) it must be fully consistent with the
braiding of $\vecu$ with other objects in comparison to the braiding of
$\vecv\vecx$ computed from Lemma~3.4.  \endproof

This says that the tensor product of a quantum covector with a quantum vector,
when treated with the correct braid statistics (i.e. as `braided covectors' and
`braided vectors'), is a braided matrix. Also,

\begin{theorem} The invariant element $\vecx\vecv\in V^*(R)\tens V(R)$ maps
under
\[V^*(R)\tens V(R){\buildrel \Psi_{V^*,V}\over\to}V(R)\tens V^*(R)\] to the
invariant element $\Psi(\vecx\vecv)=\trace\vecv\vecx \vartheta$ where $\trace$
is the ordinary matrix trace and $\vartheta^i{}_j=\tilde R^i{}_k{}^k{}_j$. This
element $\Psi(\vecx\vecv)$ is central in the algebra $V(R)\und\tens V^*(R)$.
Likewise, $\trace\vecu \vartheta$ and more generally $\trace \vecu^n\vartheta$
are invariant and central in $B(R)$.
\end{theorem}
\proof Firstly, $\Psi(\vecx\vecv)$ must be bosonic (i.e. $A$-invariant) since
$\Psi$ is covariant so it must take invariant elements to invariant elements.
Computing it from Proposition~3.2  we have $\Psi(\vecx\vecv)=\tilde
R^i{}_k{}^k{}_jv^jx_i=\trace\vartheta \vecv\vecx$. In view of the preceding
proposition we are led also to propose $\trace \vartheta\vecu$ as an invariant
element. We prove this and that $\trace \vartheta\vecu$ is
central in $B(R)$, which also implies this for its image in $V(R)\und\tens
V^*(R)$ (and similarly for higher powers of $\vecu$). The proof depends on the
theory of dual quasitriangular Hopf algebras, for in any such Hopf algebra
there is a linear functional $\vartheta:A\to k$ defined by
$\vartheta(a)=\CR(a\o\tens Sa\t)$ and obeying
$a\o\vartheta(a\t)=\vartheta(a\o)S^2a\t$ where $\Delta a=a\o\tens a\t$ (formal
sum). For proof see \cite[Appendix]{Ma:bg}, or argue by duality with a
well-known result for quasitriangular Hopf algebras. To apply this to $\vect$
we define the matrix $\vartheta=\vartheta(\vect)$ so that the above well-known
result becomes
\eqn{var1}{\vect\vartheta=\vartheta S^2\vect.}
We can now compute that $\trace \vecu^n\vartheta$ transforms to
$\trace(S\vect)\vecu^n\vect\vartheta=\trace \vecu^n\vartheta
(S^2\vect)\cdot_{\rm op}S\vect=\trace\vecu^n\vartheta S(\vect
S\vect)=\trace\vecu^n\vartheta S(1)=\trace\vecu^n\vartheta$.
Here $\cdot_{\rm op}$ denotes the reverse multiplication in $A$ and we used
that $S$ is an antialgebra map. Although it plays the role of inverse, we were
careful not to suppose that $S^2=\id$. Clearly $\vecu^n$ here can be any matrix
of generators transforming as a quantum matrix. To prove centrality let us note
two other useful properties of $\vartheta$ as defined above, namely
\eqn{var}{\vartheta_2=R\vartheta_2\tilde R,\quad \vartheta_1=\tilde
R\vartheta_1 R.}
The proof of the first of these is
$(R^{-1}\vartheta_2)^a{}_b{}^i{}_k=\CR(St^a{}_b\tens
t^i{}_j)\vartheta^j{}_k=\vartheta^i{}_j\CR(St^a{}_b\tens S^2
t^j{}_k)=\vartheta^i{}_j\CR(t^a{}_b\tens St^j{}_k)=(\vartheta_2\tilde
R)^a{}_b{}^i{}_k$ by (\ref{var1}) and invariance of $\CR$ under $S\tens S$. The
proof of the second is similar, $\CR(t^i{}_j\tens
St^a{}_b)\vartheta^j{}_k=\vartheta^i{}_j\CR(S^2 t^j{}_k\tens
St^a{}_b)=\vartheta^i{}_j R^{-1}{}^j{}_k{}^a{}_b$. We note also that
(\ref{B(R)}) implies by iteration that
\eqn{un}{R_{21}\vecu_1 R_{12}\vecu_2^n=\vecu_2 R_{21}\vecu_1
R_{12}\vecu_2^{n-1}=\cdots =\vecu_2^n R_{21}\vecu_1 R_{12}}
and applying $\trace(\ )\vartheta$ to this we have $\trace_2\vecu_1
R_{12}\vecu_2^n R_{12}^{-1}\vartheta_2=\trace_2R^{-1}_{21}\vecu_2^n
R_{21}\vecu_1 \vartheta_2$. Computing the left hand side with the aid of the
first of (\ref{var}) we have $\vecu_1\trace_2 R_{12}\vecu_2^n\vartheta_2\tilde
R_{12}=
\vecu_1\trace_2\vecu_2^n\vartheta_2\tilde R_{12}\cdot_{\rm
op_1}R_{12}=\vecu_1\trace_2\vecu_2^n\vartheta_2$ where $(\tilde
R_{12}\cdot_{\rm op_1}R_{12})^i{}_j{}^k{}_l=\tilde R^a{}_j{}^k{}_b
R^i{}_a{}^b{}_l=\delta^i_j\delta^k_l$. Similarly on the right hand side we move
$\vartheta_2$ and apply the second of (\ref{var}) (with permuted indices) to
obtain $\trace_2\tilde R_{21}\vartheta_2\vecu_2^n
R_{21}\vecu_1=(\trace_2\vartheta_2\vecu_2^n R_{21}\cdot_{\rm op_1}\tilde
R_{21})\vecu_1=\trace_2\vecu_2^n\vartheta_2\vecu_1$. \endproof

Let us note that this `quantum trace' $\trace(\ )\vartheta$ is nothing other
than a version of the abstract category theoretic trace for any braided
category with dual objects. This has been studied previously in, for example,
\cite{Ma:any} where we gave the anyonic trace as a generalization of the
super-trace. The main difference between that setting and the one above is that
previously we worked with quantum groups not dual quantum groups, and hence
with an element $(S\CR\ut)\CR\uo$ rather than $\vartheta$ above. There is also
a change from left-handed to right-handed conventions. The present form is
particularly useful because in some cases $B(R)$ is also isomorphic
to important algebras (such as the Sklyanin algebra\cite{Ma:skl}) as well as,
in a quotient, the algebra of $U_q(g)$.
In these cases the quantum trace in the theorem maps to central elements in the
algebra, a fact that is already quite well-known in these cases\cite{FRT:lie}.
Our derivation of the quantum trace as the element corresponding to
$\Psi(\vecx\vecv)$ from the point of view of braided linear algebra, is a novel
interpretation even in these cases.

Related to the invariant trace, should be a braided determinant. The braided
determinant $\BDET (\vecu)$ should be bosonic (i.e. have trivial braid
statistics) central and group-like according to $\und\Delta \BDET (\vecu)=\BDET
(\vecu)\tens \BDET (\vecu)$, so that $\BDET (\vecu\vecu')=\BDET (\vecu)\BDET
(\vecu')$ when $\vecu,\vecu'$ are treated with the braid statistics from
Lemma~3.4. In addition, we can expect that the braided-determinant of a rank
one quantum matrix should be zero, i.e.
\eqn{BDET}{\BDET (\vecv\vecx)=0.}
A general treatment of this topic must surely await a treatment of braided
exterior algebras, but we shall at least see these properties in some explicit
examples in the next section.

Also, now that we have introduced our braided matrices $B(R)$ we can use it to
act on covectors and vectors, as well as itself, in a fully covariant way with
respect to the hidden dual quantum group symmetry $A$. Thus we have analogs of
Examples~2.4,2.5 and Lemma~2.6 as follows.

\begin{propos} The assignment $\vecx'=\vecx\vecu$ makes $V^*(R)$ into a right
braided $B(R)$-comodule algebra, i.e. gives a (covariant) algebra homomorphism
$B(R)\to V^*(R)\und\tens B(R)$. Thus, provided $\vecx,\vecu$ are treated with
the
correct braid statistics, $\vecx'$ is also a realization of $V^*(R)$.
\end{propos}
\proof We use the braid statistics
$\vecu_1\vecx_2:=\Psi(\vecu_1\vecx_2)=\vecx_2R^{-1}\vecu_1R$ from Lemma~3.4,
and associativity of the braided tensor product algebra. Thus
$\vecx'_1\vecx'_2=\vecx_1\vecu_1\vecx_2\vecu_2:=\vecx_1\vecx_2
R^{-1}_{12}\vecu_1R_{12}\vecu_2=\lambda\vecx_2\vecx_1\vecu_1R_{12}\vecu_2$ from
the relations in $V^*(R)$. Meanwhile, we also have $\lambda
\vecx_2'\vecx_1'R=\lambda \vecx_2\vecu_2\vecx_1\vecu_1 R_{12} :=$ $
\lambda\vecx_2\vecx_1R_{21}^{-1}\vecu_2 R_{21}\vecu_1 R_{12}$ using the braid
statistics again (with indices permuted). These expressions are equal after
using (\ref{B(R)}). Hence $\vecx'$ is also a realization of $V^*(R)$.
The construction is manifestly covariant under the background dual quantum
group $A$.
\endproof

\begin{propos} Let $B$ be the braided group obtained by quotienting $B(R)$,
with braided-antipode $\und S$. We write $\vecu^{-1}=\und S\vecu$. The
assignment $\vecv'=\vecu^{-1}\vecv$ makes $V(R)$ into a right braided
$B$-comodule algebra, i.e. gives a (covariant) algebra homomorphism $B\to
V(R)\und\tens B$. Thus, provided $\vecx,\vecu$ are treated with the correct
braid statistics, $\vecv'$ is also a realization of $V(R)$.
\end{propos}
\proof Firstly, it is necessary to quotient $B(R)$ by suitable `braided
determinant-type' relations (or invert suitable elements) such that the braided
antipode $\und S$ exists and is compatible with the braiding (this is possible
whenever it is possible for the corresponding ordinary dual quantum group).
This is the content of our regularity assumption on $R$. We denote the
resulting braided matrix group\cite{Ma:exa} by $B$ and the result of the
braided antipode by $\vecu^{-1}$. The axioms
for it are the same as the usual ones (but with respect to the braided
comultiplication), so  $\vecu^{-1}\vecu=1=\vecu\vecu^{-1}$. Most importantly
for us, this map $\und S$ (like all the braided group maps) is covariant so
that it commutes with $\Psi$. This means that the braid statistics of
$\vecu^{-1}$ with $\vecv$ are
read off from Proposition~3.2 or Lemma~3.4 for $\vecu^{-1}$ transforming as a
quantum matrix (in place of $\vecu$ there). These statistics are essential
because the meaning of $\vecu^{-1}\vecv$ is precisely
$\vecu^{-1}\vecv:=\Psi(\vecu^{-1}\vecv)$ by definition as an element of the
braided tensor product algebra $V(R)\und\tens B$. We write $\vecu^{-1}\vecv$
with $\vecu^{-1}$ on the left for convenience with regard to its matrix
structure, but it officially belongs on the right of the $\vecv$ after using
the cross relations. In practice, it is convenient to write the statistics
relations in the implicit form $R^{-1}\vecu^{-1}_1
R\vecv_2:=\vecv_2\vecu_1^{-1}$ (i.e. $\Psi(R^{-1}\vecu^{-1}_1
R\vecv_2)=\vecv_2\vecu_1^{-1}$) from Lemma~3.4. Then
$\vecv_1'\vecv_2'=\vecu_1^{-1}\vecv_1\vecu_2^{-1}\vecv_2
=:\vecu_1^{-1}R_{21}^{-1}\vecu_2^{-1}R_{21}\vecv_1\vecv_2=R_{12}
\vecu_2^{-1}R_{12}^{-1}\vecu_1^{-1}\vecv_1\vecv_2=
\lambda R_{12}\vecu_2^{-1}R_{12}^{-1}\vecu_1^{-1}R_{12}\vecv_2\vecv_1:=\lambda
R_{12}\vecu_2^{-1}\vecv_2\vecu_1^{-1}\vecv_1=\lambda R\vecv_2'\vecv_1'$. We
used the relations (\ref{B(R)}) and the defining relations of $V(R)$, as well
as the statistics relations as explained. Thus the $\vecv'$ also realise
$V(R)$, and in a manifestly covariant way. \endproof

\begin{theorem} $B$ obtained from $B(R)$ acts on itself in the sense that the
assignment $\vecu''={\vecu'}^{-1}\vecu\vecu'$ makes $B$ into a right braided
$B$-comodule algebra, where $\vecu'$ denotes the second (coacting) copy of $B$.
Thus, (provided $\vecu,\vecu'$ are treated with the
correct braid statistics) $\vecu''$ is also a realization of $B$ and so
provides a (covariant) algebra homomorphism $B\to B\und\tens B$. We call it the
{\em braided adjoint coaction} of $B$ on itself.
\end{theorem}
\proof As in the previous proposition, the expression ${\vecu'}^{-1}
\vecu\vecu'$ means $\Psi({\vecu'}^{-1}\vecu)\vecu'$ where the statistics
relations between ${\vecu'}^{-1}$ and $\vecu$ must be used if we want to
exhibit this as an element of $B\und\tens B$ with the $\vecu'$ parts living in
the second factor of $B$. It is essential to keep these statistics in mind.
Again, we need to be careful not to confuse $\Psi$ with $\Psi^{-1}$. In the
present computation there is no danger of this because all elements living in
the second (coacting) factor of $B\und\tens B$ are labeled with a prime.
Thus $g'h$ always means $\Psi(g'h)$ for $h$ in the first factor and $g'$ in the
second factor of the resulting expression. The prime means there is no danger
of confusion with $hg'=h\tens g'$ in $B\und\tens B$, so we will suppress the
$:=$ distinction (we could have used a similar device in the proofs of the
preceding two propositions). Thus we  just work with the associative algebra
$B\und\tens B$ generated by the relations of $B$ on primed and unprimed
variables from (\ref{B(R)}) and the cross relations
\eqn{cro}{R^{-1}\vecu'_1 R\vecu_2=\vecu_2R^{-1}\vecu'_1R,\quad }
from Lemma~3.4. Since the braiding is functorial, the same cross relations hold
for ${\vecu'}^{-1}$ in place of $\vecu'$, which is the form that we will use.
Then \align{&&\nqquad
R_{21}{\vecu'}'_1R_{12}{\vecu'}'_2=R_{21}{\vecu'}_1^{-1}\vecu_1(\vecu'_1R_{12}
{\vecu'}_2^{-1})\vecu_2\vecu'_2\\
&&\nqquad
\quad=R_{21}{\vecu'}_1^{-1}(\vecu_1R_{21}^{-1}{\vecu'}_2^{-1}R_{21})
(\vecu'_1R_{12}\vecu_2)\vecu'_2=(R_{21}{\vecu'}_1^{-1}R_{21}^{-1}{\vecu'}_2^{-1})
(R_{21}\vecu_1R_{12}\vecu_2 R_{12}^{-1})\vecu'_1R_{12}\vecu'_2\\
&&\nqquad
\quad={\vecu'}_2^{-1}(R_{12}^{-1}{\vecu'}_1^{-1}R_{12}\vecu_2)R_{21}\vecu_1
(\vecu'_1R_{12}\vecu'_2)={\vecu'}_2^{-1}\vecu_2 R^{-1}_{12}{\vecu'}_1^{-1}
R_{12}(R_{21}\vecu_1R_{21}^{-1}\vecu'_2R_{21})
\vecu'_1R_{12}\\
&&\nqquad \quad={\vecu'}_2^{-1}\vecu_2 (R^{-1}_{12} {\vecu'}_1^{-1}
R_{12}\vecu'_2 R_{21} )\vecu_1\vecu'_1R_{12} = {\vecu'}_2^{-1}\vecu_2
\vecu'_2R_{21}{\vecu'}_1^{-1}\vecu_1\vecu'_1R_{12}=\vecu''_2R_{21}
\vecu''_1R_{12}.}
We used only the relations for the braided tensor product $B\und\tens B$ in the
form described, applied in each expression to the parts in parentheses to
obtain the next expression. Thus $\vecu''$ is a (manifestly covariant)
realization of $B$. \endproof

The abstract picture behind the theorem is as follows. Just as any dual quantum
group coacts on itself by the adjoint coaction, so every Hopf algebra $B$ in a
braided category coacts on itself by the braided adjoint coaction. In the
quantum group case we saw, as in Lemma~2.6 that this coaction of a dual quantum
group on itself does {\em not} respect its own algebra structure (unless the
dual quantum group is commutative). The same is true in general
in the braided setting, the braided adjoint coaction of $B$ on itself  does not
in general respect the algebra of $B$ unless $B$ is `braided-commutative' in a
certain precise sense. See the appendix. But the braided groups $B$ obtained by
transmutation {\em are} commutative in precisely the right sense, see
\cite{Ma:bos} where we show this (in the form of the braided adjoint actions
and braided-cocommutativity rather than coactions as here). The $B$ in
Theorem~3.9 is just
of this  type (it is formally the transmutation of the dual quantum group $A$),
and this is the abstract reason behind the result. Thus, dual quantum groups
are not full covariant under their own adjoint coaction, but the process of
transmutation turns the dual quantum group into (the functions on) an actual
group (a braided-commutative ring
 of functions, like the super-commutative ring of functions on a super-group).
The non-commutativity of the dual quantum group is placed now in the braided
category (i.e. in the braid statistics) and after allowing for these, the
resulting object  behaves like a classical (not quantum) group. Because of
this, it acts on itself by conjugation just as ordinary (not quantum) groups
do. This is the rationale (apart from covariance) behind the introduction of
braided groups in \cite{Ma:eul}\cite{Ma:bg}\cite{Ma:exa}. Theorem~3.9 confirms
the usefulness of this picture.

\section{Examples}

In this section we develop several examples of the general covariant-quantum
(braided) linear algebra above. Before describing these, we need to make a note
about the normalization of the $R$ matrices. In the above we have assumed that
$R$ is regular in the sense that it is the restriction to the generators
$\vect$ of a dual quasitriangular Hopf algebra $A$ obtained from $A(R)$. In
general such dual-quasitriangular structures cannot be rescaled (the axioms are
not linear). However, if we concentrate on the bialgebra $A(R)$ rather than any
special quotient $A$, then we are free to rescale.
This is because $A(R)$ is a quadratic algebra (in particular, with homogeneous
relations) so that every element has a well-defined degree (the number of
generators making up the element). Moreover, the matrix form of the
comultiplication means that each factor of $\Delta a$ has the same degree as
$a$. Hence if $\CR$ is a dual quasitriangular structure then so is $\CR'(a\tens
b)=\lambda^{\deg(a)\deg(b)}\CR(a\tens b)$. This is the dual quasitriangular
structure corresponding to the rescaled matrix $R'=\lambda R$.

Thus, at the general bialgebra level we are free to rescale $R$. We have given
direct matrix proofs of all the main results above and it is clear that these
results hold even at this general bialgebra level (where we do not worry about
the existence of $A$ as an honest dual quantum group) provided $R^{-1},\tilde
R,\vartheta$ exist with various matrix properties. This is the level at which
we will work in the present section. We note that the relations of $B(R)$ and
its statistics, as well as the statistics between $\vecu$ and $\vecx,\vecv$ are
in any case independent of the normalization of $R$. Meanwhile, the relations
of $V(R),V^*(R)$ already have a specific parameter $\lambda$ to accommodate
different
normalizations, so that only the statistics relations of $V(R),V^*(R)$ are
affected. These {\em do} depend on the normalization of $R$, but let us note
that $V(R),V^*(R)$ are again quadratic  algebras (with homogeneous relations).
Clearly, if $\Psi$ is a braiding on them then $\Psi'(x\tens
y)=\lambda^{\deg(x)\deg(y)}\Psi(x\tens y)$ on homogeneous elements $x,y$ is
also a braiding.  This is all that happens when we change the normalization of
$R$, and in the examples below we can exploit it to put the braidings on
$V(R),V^*(R)$ in the simplest form.

Our examples are as follows. We begin with the obligatory example of the
standard $SL_q(2)$ $R$-matrix. It demonstrates the features of the general
standard $R$ matrices also. We next give its two-parameter variant, the
$GL_{p,q}$ R-matrix studied in \cite{DMMZ:non}\cite{Tak:two} and elsewhere.
This is followed by the non-standard variant related to the Alexander-Conway
knot polynomial (where the braided linear algebra reduces to super-linear
algebra as $q\mapsto 1$). Finally, we study an $R$-matrix
connected with the 8-vertex model. Its dual quantum group $A(R)$ was recently
studied in \cite{GJL:mod}. Each of these examples demonstrates a different
aspect of the theory. All the examples have $4\times 4$ $R$-matrices and we
denote the four braided matrix generators by $\vecu=\pmatrix{a&b\cr c&d}$. We
denote the two  covector generators by $\vecx=(x\ y)$ and the two vector
generators by $\vecv=\pmatrix{v\cr w}$. Some of the computations have been done
with the assistance of the computer package REDUCE. For simplicity we state the
results over a field of characteristic zero, such as $k=\C$.

\subsection{Standard R-Matrix} Here we note how the constructions of braided
(i.e. covariant-quantum) linear algebra look for the standard solution of the
QYBE corresponding to the dual quantum group $SL_q(2)$ and to the Jones knot
polynomial. This provides orientation for the non-standard examples that
follow. We take the normalization that gives (with $\lambda=1$) the standard
(not fermionic) quantum planes for the vectors and covectors. This is
\[ R(q)=\pmatrix{1&0&0&0\cr 0&q^{-1}&1-q^{-2}&0\cr 0&0&q^{-1}&0\cr
0&0&0&1},\quad\tilde R=\pmatrix{1&0&0&0\cr 0&q&q^{-2}-1&0\cr 0&0&q&0\cr
0&0&0&1},\quad \vartheta=\pmatrix{q^{-2}&0\cr 0&1}\]
and $R^{-1}=R(q^{-1})$. The braided matrices $B(R)=BM_q(2)$ were given in
\cite{Ma:exa} along with their statistics $\Psi$. We do not repeat its details
here, but note only that all the algebra relations and braiding depend on $q^2$
and not directly on $q$ itself. This is also true for the invariant trace
element $\trace\vecu\vartheta=q^{-2}a+d$  and for the braided determinant
\eqn{slbdet}{\BDET (\vecu)=ad-q^2cb}
found in \cite{Ma:exa}. In \cite{Ma:skl} we show that the algebra $BM_q(2)$
(with some elements taken invertible) is isomorphic to the degenerate Sklyanin
algebra and $\trace \vecu\vartheta$ and $\BDET (\vecu)$ become its two
Casimirs. After setting $\BDET (\vecu)=1$ we also obtain $BSL_q(2)\isom
U_q(sl(2))$ as an algebra and its quadratic Casimir as usual.

The braided algebra of covectors $V^*(R)$ is
\eqn{slcovec}{xy=q^{-1}yx}
\[\Psi(x\tens x)=x\tens x,\quad \Psi(x\tens y)=y\tens x q^{-1}\]
\eqn{slcovecpsi}{ \Psi(y\tens x)=x\tens yq^{-1}+(1-q^{-2})y\tens x,\quad
\Psi(y\tens y)=y\tens y}
Recall that there are various notations for the braiding. For example, written
as the statistics relations (the cross relations in the algebra
$V^*(R)\und\tens V^*(R)$) it is $x'x=xx'$, $x'y=q^{-1}yx'$,
$y'x=xy'q^{-1}+(1-q^{-2})yx'$ and $y'y=yy'$ as explained above. We see that
over $\C$ the algebra is that of the standard quantum plane $\C^2_{q^{-1}}$.
The vectors $V(R)$ are similar, namely
\eqn{slvec}{vw=qwv}
\[\Psi(v\tens v)=v\tens v,\quad \Psi(v\tens w)=w\tens vq^{-1}+(1-q^{-2})v\tens
w\]
\eqn{slvecpsi}{  \Psi(w\tens v)=v\tens w q^{-1},\quad \Psi(w\tens w)=w\tens w.}
This is isomorphic to the covectors with $w,v$ in the role of $x,y$ (note the
reversal).

The cross relations in $V(R)\und\tens V^*(R)$ (i.e. the braidings $\Psi(x\tens
v)$ etc) are
\eqn{slcro}{xv:=vx,\quad xw:=qwx,\quad yv:=qvy,\quad yw:=wy+(q^{-2}-1)vx}
These relations together with the algebra relations
(\ref{slcovec})(\ref{slvec}) give the algebra $V(R)\und\tens V^*(R)$. Our
general theory says that $BM_q(2)$ (i.e. the degenerate Sklyanin algebra) is
realized in this braided tensor product. We compute $\BDET (\vecu)$ in this
realization, i.e. $\BDET (\vecv\vecx)$ as an element of $V(R)\und\tens V^*(R)$.
We have $a=vx, b=vy$ etc so that $\BDET
(\vecv\vecx)=vxwy-q^2wxvy:=qvwxy-q^2wvxy=q^2wvxy-q^2wvxy=0$ using the relations
in $V(R)\und\tens V^*(R)$. Thus the $\BDET $ on $BM_q(2)$, in addition to being
(central) bosonic and group-like, vanishes on rank-one matrices as in
(\ref{BDET}).

The braiding $\Psi(v\tens x)$ etc in the other direction (the cross relations
in $V^*(R)\und\tens V(R)$) are similar but different and easily computed in the
same way, as are the braidings $\Psi(\vecu\tens \vecv)$ and
$\Psi(\vecv\tens\vecu)$ from Lemma~3.4. With these braid statistics, one can
verify the action of $BSL_q(2)$ on the braided vectors and covectors, and on
itself as in Theorem~3.9. In another normalization we have `fermionic'
versions of the above. Some similar results apply for all the standard
$R$-matrixes corresponding to simple Lie algebras $g$.

\subsection{2-Parameter Solution} Here we note the results for the 2-parameter
solution of the QYBE leading to the quantum matrices $M_{p,q}(2)$ and dual
quantum group $GL_{p,q}(2)$ after inverting some elements. This dual quantum
group has been studied by several authors, notably
\cite{DMMZ:non}\cite{Tak:two}. In the normalization that we use we have,
\[ R(p,q)=\pmatrix{1&0&0&0\cr 0&p&1-pq&0\cr 0&0&q&0\cr 0&0&0&1},\quad\tilde
R=\pmatrix{1&0&0&0\cr 0&p^{-1}&pq-1&0\cr 0&0&q^{-1}&0\cr 0&0&0&1},\quad
\vartheta=\pmatrix{pq&0\cr 0&1}\]
and $R^{-1}=R(p^{-1},q^{-1})$. Our first task is to compute the braided
matrices $B(R)=BM_{p,q}(2)$, say. Using the relations (\ref{B(R)}) one finds
\[ BM_{p,q}(2)=BM_{(pq)^{-\h}}(2),\quad \BDET (\vecu)=ad-p^{-1}q^{-1}cb,\quad
\trace\vecu\vartheta= pqa+d\]
Thus, although the dual quantum group for this solution is different from the
$SL_q(2)$ or $GL_q(2)$ case above, the braided group comes out the same. Recall
above that $BM_q(2)$, its invariant trace element and braided-determinant etc
depended only on $q^{2}$ (not on $q$ itself). That $q^{2}$ is factorizing now
into $p^{-1},q^{-1}$. Note also that whereas the ordinary quantum determinant
is not central\cite{DMMZ:non}, the braided determinant is central in
$BM_{p,q}(2)$.

The braided algebra of covectors $V^*(R)$ is
\eqn{pqcovec}{xy=pyx}
\[\Psi(x\tens x)=x\tens x,\quad \Psi(x\tens y)=y\tens x p\]
\eqn{pqcovecpsi}{ \Psi(y\tens x)=x\tens yq+(1-pq)y\tens x,\quad \Psi(y\tens
y)=y\tens y}
Let us call this braided algebra $\C^2_{p,q}$ if we work over $\C$. The $p$ is
the quantum parameter controlling the non-commutativity of the algebra, and the
additional $q$ (along with $p$) is a parameter appearing in the statistics
relations. The vectors $V(R)$ are similar, namely
\eqn{pqvec}{vw=q^{-1}wv}
\[\Psi(v\tens v)=v\tens v,\quad \Psi(v\tens w)=w\tens vp+(1-pq)v\tens w\]
\eqn{pqvecpsi}{  \Psi(w\tens v)=v\tens w q,\quad \Psi(w\tens w)=w\tens w.}
Thus, as braided algebras we have that $w,v$ in place of $x,y$ (note the
reversal) generate $\C^2_{q,p}.$

The cross relations in $V(R)\und\tens V^*(R)$  are
\eqn{pqcro}{xv:=vx,\quad xw:=p^{-1}wx,\quad yv:=q^{-1}vy,\quad yw:=wy+(pq-1)vx}
These relations together with the algebra relations
(\ref{pqcovec})(\ref{pqvec}) give the algebra $V(R)\und\tens V^*(R)$. The other
statistics relations can be obtained similarly. As before, these statistics
relations for $V(R)\und\tens V^*(R)$ and a computation similar to the preceding
example gives that $\BDET (\vecv\vecx)=0$ as it should.

\subsection{Alexander-Conway Solution} Here we mention the analogous results
for the non-standard $R$-matrix studied by various authors and known to be
connected with the Alexander-Conway polynomial. A recent work is
\cite{MaPla:uni} (where the full quantum group structure, including the
quasitriangular structure, is found). The $R$-matrix in one suitable
normalization is
\[ R(q)=\pmatrix{1&0&0&0\cr 0&q^{-1}&1-q^{-2}&0\cr 0&0&q^{-1}&0\cr
0&0&0&-q^{-2}},\quad\tilde R=\pmatrix{1&0&0&0\cr 0&q&q^2-1&0\cr 0&0&q&0\cr
0&0&0&-q^2},\quad \vartheta=\pmatrix{q^2&0\cr 0&-q^2}\]
and $R^{-1}=R(q^{-1})$.  The corresponding braided matrices $B(R)$ (as well as
the invariant trace element in this case) were computed in \cite{Ma:exa}. We
assume that $q^2\ne -1$.

The covectors and vectors as braided algebras are
\eqn{acvec}{xy=q^{-1}yx,\quad y^2=0,\qquad vw=qwv,\quad w^2=0}
\eqn{acpsi}{\Psi(y\tens y)=-q^{-2}y\tens y,\quad \Psi(w\tens w)=-q^{-2}w\tens
w}
with the other $\Psi$ as in (\ref{slcovecpsi})(\ref{slvecpsi}). The statistical
cross relations in $V(R)\und\tens V^*(R)$ are also modified to
\eqn{accro}{xv:=vx,\quad xw:=qwx,\quad yv:=qvy,\quad yw:=-q^{2}wy+(q^2-1)vx.}
In these equations the main difference from the standard example in Section~4.1
is that $y$ and $w$ become $q$-deformations of fermionic variables in terms of
their various relations and statistics (as do the elements $b,c$ of the braided
matrices). In the limit $q\mapsto 1$, the braided matrices $B(R)$ become the
super-matrices $M_{1|1}$\cite{Ma:exa}, and $V(R),V^*(R)$ become $1|1$-super
planes. In another choice of normalization, it is the $x,v$ rather than the
$y,w$ that become `fermionic'. Note that a connection between this Yang-Baxter
matrix and super-symmetry is well established in a physical way in
\cite{KauSal:fre}, but here we see the connection at the level of elementary
$q$-deformed super-linear algebra.

\subsection{8-Vertex Solution} Here we give the details for a less-well known
$R$-matrix related to the 8-Vertex model in statistical mechanics. Its
bialgebra $A(R)$ was studied recently in \cite{GJL:mod} and is non-commutative.
The $R$-matrix for our purposes is
\[ R(q)=(q+1)^{-1}\pmatrix{1&0&0&nq\cr 0&m&q&0\cr 0&q&m&0\cr nq&0&0&1},\qquad
\tilde R=R^{-1}=R(-q),\quad \vartheta=\pmatrix{1&0\cr 0&1}\]
where $q^2\ne 1$ and $m^2=1=n^2$. Our first goal is to compute $B(R)$ from
(\ref{B(R)}). After some computation one finds
\eqn{8verB(R)}{\{a,b,c,d\}\ {\rm commute},\qquad b^2=c^2,\ a^2=d^2,\ ac=mnbd,\
cd=mnba}
so that the braided matrices, in addition to being `braided commutative' are
actually commutative! The braiding from Proposition~3.2 is however,
non-trivial. To describe it, it is convenient to choose new generators
\[ D=d-a,\qquad B=b-mnc,\qquad C_1=d+a,\qquad C_2=b+mnc.\]
In these variables, the relations of $B(R)$ are
\[ BC_i=0,\quad DC_i=0;\qquad  B(R)=k[B,D]\oplus k[C_1,C_2]\]
where we mean that $B(R)$ is generated by polynomials in $B,D$ and by
polynomials in $C_1,C_2$, and apart from the identity, the product of an
element in one polynomial algebra with an element from the other is zero. This
means that (apart from the identity element, which is common to both), the
algebra splits as a direct sum.
The underlying variety can be thought of as the union of two planes, one at
$C_1=C_2=0$ and the other at $B=D=0$. The element $C_1$ is the invariant trace
element and so is necessarily bosonic (i.e. has trivial braiding with
everything else), but it turns out that $C_2$ is also bosonic. The remaining
statistics between $B,D$ take the form
\[ \Psi(B\tens B)=\alpha B\tens B+\beta mn\, D\tens D,\quad \Psi(B\tens
D)=\alpha D\tens B+\beta B\tens D\]
\eqn{8verpsi}{ \Psi(D\tens B)=\alpha B\tens D+\beta D\tens B,\quad \Psi(D\tens
D)=\alpha D\tens D+\beta mn\, B\tens B}
\eqn{alphabeta}{\alpha={q^4+6q^2+1\over (q^2-1)^2},\qquad
\beta={4q(q^2+1)\over(q^2-1)^2}.}
In fact, the matrix describing these braid statistics is of the same type as
$R$ itself, with new values of parameters $n'=nm,q'=\beta/\alpha$ and  $m'=1$.
Using the relation $\alpha^2-\beta^2=1$ between the rational functions
$\alpha,\beta$, it is easy to see that $D^2-mnB^2$ and $C_1^2-mnC^2_2$ are
bosonic. The matrix comultiplication $\und\Delta$ means that these elements are
not themselves group-like but noting that $\alpha+\beta=({q+1\over q-1})^4$,
one finds that the combination
\eqn{8verbdet}{ \BDET (\vecu)={\scriptstyle{1\over 4}}(C_1^2-mn
C_2^2-({q+1\over
q-1})^2(D^2-mnB^2))={q^2+1\over(q-1)^2}(ad-bc)-{2q\over(q-1)^2}(a^2-mnb^2)}
is group-like. Recall that $\und\Delta$ extends to products as an algebra
homomorphism to the braided tensor product. This (\ref{8verbdet}) is the
braided-determinant for $B(R)$. If we set $\BDET (\vecu)=1$ we obtain a braided
group with braided-antipode
\eqn{8verant}{\und S\pmatrix{a&b\cr c&d}=(q-1)^{-2}\pmatrix{(q^2+1)d-2qa&
-(q^2+1)b+2mnqc\cr -(q^2+1)c+2mnqb&(q^2+1)a-2qd}.}
Thus completes our description of the braided-matrices and braided group for
this $R$-matrix. Its classical limit is at $q=0$, with another classical limit
(with the same braided matrices and braided group) at $q=\infty$ in a suitable
sense.

The covectors $V^*(R)$ and vectors $V(R)$ for the above normalization and for
$q\ne 0$ are given by
\eqn{8vervec}{xy=myx,\quad x^2=ny^2,\qquad vw=mwv,\quad v^2=nw^2}
\[\Psi(x\tens x)=x\tens x+nqy\tens y,\quad \Psi(x\tens y)=y\tens x m+ qx\tens
y\]
\eqn{8vervecpsi}{\Psi(y\tens x)=x\tens y m+qy\tens x,\quad \Psi(y\tens
y)=y\tens y+nq x\tens x}
with braiding on $V(R)$ given by the same formulae with $v,w$ in the role of
$x,y$. We have suppressed an overall factor $(q+1)^{-1}$ on the right hand
sides.  The statistics relations between these two braided algebras, i.e. the
cross relations in $V(R)\und\tens V^*(R)$ are
\eqn{8vercro}{xv:=vx-qwy,\quad xw:=mwx-nqvy,\quad yv:=mvy-nqwx,\quad
yw:=wy-qvx}
with an overall factor $(1-q)^{-1}$ suppressed on the right hand sides. Using
these and the relations (\ref{8vervec}) in each algebra we can easily verify
that $a,b,c,d$ when realized in $V(R)\und\tens V^*(R)$ really are mutually
commutative as they must be by Proposition~3.5. For example,
$ab=vxvy:=v^2xy-qvwy^2=v^2myx-qnvwx^2=:vyvx=ba$ etc. Also, we can compute
$D^2/2=a^2-ad=vxvx-wyvx:=v^2x^2-qvwyx-mwvyx+nqw^2x^2=(1+q)(v^2x^2-wvxy)$ while
similarly
$B^2/2=b^2-mnbc=vyvy-mnvywx:=v^2y^2m-nqvwxy-mnvwyx+mnqv^2x^2=mnD^2/2$. Thus
$D^2-mnB^2=0$ in this realization of $B(R)$. Likewise, one computes that
$C_1^2-mn C_2^2=0$ in this realization. Hence we have $\BDET (\vecv\vecx)=0$ as
it should on our braided rank-one matrices.

\section{Transmutation by Sewing}

In Section~2 we have described quantum linear algebra and in Section~3
developed a covariant braided version based on the braided matrixes $B(R)$
acting rather than the dual quantum group $A(R)$. By way of concluding remarks
we now study further the process of transmutation that relates the two. The
situation here for general dual quantum groups is given in
\cite{Ma:eul}\cite{Ma:bg}, but we want to note the form that it takes in the
matrix case.

Firstly, we recall that a dual quantum group $A$ (in the strict sense) comes
equipped with a dual quasitriangular structure $\CR:A\tens A\to k$ obeying some
obvious axioms dual to those of Drinfeld for a quasitriangular Hopf algebra.
For $A(R)$ it is given by the matrix $R$ in the generators and extended in such
as way that $\CR((\ )\tens t^i{}_j)$ is a matrix representation of $A$, and
$\CR(t^i{}_j\tens(\ ))$ is a matrix anti-representation. We showed this in some
form in \cite[Sec. 3.2.3]{Ma:qua} and called it the bimultiplicativity property
of $\CR$ in \cite[Sec. 4.1]{Ma:pro}. See also \cite{LarTow:two} and others.

Now according to \cite{Ma:bg} the structure of $B(R)$ can be realised in the
linear space of $A(R)$ but with a modified product, which we will denote
explicitly by $\und\cdot$ to distinguish it. It is not necessary here for
$A(R)$ itself to be a Hopf algebra, as long as $R$ is regular so that $A(R)$
has a quotient which becomes a dual quasitriangular Hopf algebra $A$. We
transmute with respect to the bialgebra map $A(R)\to A$. In our case it comes
out from \cite{Ma:bg} as
\eqn{trans}{u^i{}_j=t^i{}_j,\quad u^i{}_j\und\cdot
u^k{}_l=t^a{}_bt^d{}_lR^i{}_a{}^c{}_d \tilde R^b{}_j{}^k{}_c \qquad({\rm i.e.}\
\ \vecu_1\und\cdot R\vecu_2=R\vect_1\vect_2).}
Thus, the generators can identified but not their products. This is why the
$\vecu$ transform in the same way as the $\vect$ under the quantum adjoint
coaction, but only the $\und\cdot$ multiplication is covariant. What does
$\und\cdot$ look like on general elements, viewed as a modified multiplication
on $A(R)$?

To explain this we define the `partition function' $Z_R({}^A_D\quad{\ }^B_C)$
as in \cite[Sec. 5.2.1]{Ma:qua} by
\align{ Z_R({}^A_D\quad {\ }^B_C)=
&&R^{a_1}{}_{m_{11}}{}^{b_1}{}_{n_{11}}\ \,
R^{m_{11}}{}_{m_{12}}{}^{b_2}{}_{n_{21}}\ \
\cdots \ R^{m_{1N-1}}{}_{c_{1}}{}^{b_N}{}_{n_{N1}}\\
&&R^{a_2}{}_{m_{21}}{}^{n_{11}}{}_{n_{12}}R^{m_{21}}{}_{m_{22}}
{}^{n_{21}}{}_{n_{22}}\\
&&\qquad\ \vdots\qquad\qquad\qquad\qquad\qquad\qquad\quad\vdots\\
&&R^{a_M}{}_{m_{M1}}{}^{n_{1M-1}}{}_{d_1} \ \ \cdots\quad \ \, \cdots\quad
R^{m_{MN-1}}{}_{c_{M}}{}^{n_{NM-1}}{}_{d_N}}
where the $A=(a_1,a_2,\cdots,a_M)$ etc are multi-indices (arranged
consequtively following any marked orientation of the edge of the lattice). We
also write $t^I{}_J=t^{i_1}{}_{j_1}t^{i_2}{}_{j_2}\cdots t^{i_M}{}_{j_M}$ etc
as a typical element of $A(R)$. The general transmutation formula in
\cite{Ma:bg} involves computing such expressions as $\CR(t^I{}_J\tens
t^K{}_L),\CR(t^I{}_J\tens S t^K{}_L)$. Using the bimultiplicativity property of
$\CR$ explained above, we can factorise such expressions into products of $R$
and $\tilde R$ respectively. Computing in this way, we obtain
\eqn{hol}{t^I{}_J\und\cdot t^K{}_L=t^A{}_B t^D{}_L \, Z_R({}^I_D\quad {\
}^C_A)\, Z_{\tilde R}({}^B_C\quad {\ }^K_J).}

We gave a similar `partition function' description of the category-theoretic
rank or `quantum dimension' of $A(R)$ in \cite{Ma:eul}. The present expression
suggests a possible interpretation of the transmutation of the usual $A(R)$ to
the braided matrices $B(R)$ in terms of a statistical transfer matrix with the
input and output states appearing on the boundary of the lattice. This is a
little like the definition of vertex operators in string field theory, as is
perhaps the factorization into $Z_R$ and $Z_{\tilde R}$. Such a physical
interpretation is an interesting direction for further work.

\appendix
\section{Diagrammatic Proof of Braided Adjoint Coaction}

In this section we develop some of the abstract picture underlying Theorem~3.9.
We have given a direct matrix proof in the text but mentioned that the
underlying reason why it works is that $B(R)$, unlike $A(R)$, is a
(braided)-commutative object in a certain sense, much as a super-group is
super-commutative. The general setting for developing this remark is that of
braided monoidal categories and allows us to give a diagrammatic proof of the
result for any braided group (not just of matrix type). Since the tensor
products here will always be braided, we will write simply $\tens$ rather than
any special notation such as $\und\tens$. For a formal treatment of braided
monoidal categories, see \cite{JoyStr:bra}.

There is a standard diagrammatic notation for working with structures in
braided categories, which we will use here also. We write all morphisms
pointing downwards and write $\Psi,\Psi^{-1}$ as braids,
$\Psi=\searrow,\Psi^{-1}=\swarrow$. Other morphisms, such as the multiplication
$B\tens B{\buildrel\cdot\over\to}B$ of an algebra $B$ living in the category,
are written as vertices with inputs and outputs according to the valency of the
map. The functoriality of $\Psi,\Psi^{-1}$ means that we can translate these
vertices through the braid crossings (without cutting any paths). For example,
using this notation, it is easy to see that if $B$ is an algebra in the
category then the multiplication on $B\tens B$ defined with the braid
statistics $\Psi$  is associative. We will use such diagrammatic notation
freely below. For details, and for the axioms of Hopf algebras in braided
categories written out in this way, we refer to \cite{Ma:tra}\cite{Ma:bos}
where the notation is used extensively. The braided-comultiplication
$\Delta:B\to B\tens B$ and the braided-antipode $S:B\to B$ of a Hopf algebra in
the braided category, are of course required to be morphisms (and so
represented by 3- and 2-vertices). Again, since all structures are braided, we
do not explicitly underline them.

In this notation, a braided group means a pair $(B,\CO)$ where $B$ is a Hopf
algebra in the braided category and $\CO$ is a class of right $B$-comodules
(also living in the braided category) such that\cite{Ma:bg}

\medskip
\unitlength=1.05mm
\special{em:linewidth 0.4pt}
\linethickness{0.4pt}
\begin{picture}(93.00,22.00)
\put(47.00,22.00){\makebox(0,0)[cc]{$V\ \otimes\ B$}}
\put(83.00,22.00){\makebox(0,0)[cc]{$V\ \otimes\ B$}}
\put(47.00,2.00){\makebox(0,0)[cc]{$V\ \otimes\ B$}}
\put(83.00,2.00){\makebox(0,0)[cc]{$V\ \otimes\ B$}}
\put(39.00,15.00){\makebox(0,0)[cc]{$\beta$}}
\put(56.00,6.00){\makebox(0,0)[cc]{$\cdot$}}
\put(75.00,16.00){\makebox(0,0)[cc]{$\beta$}}
\put(93.00,6.00){\makebox(0,0)[cc]{$\cdot$}}
\put(65.00,13.00){\makebox(0,0)[cc]{$=$}}
\end{picture}

\noindent for all $(V,\beta)$ in $\CO$. One says that $B$ is {\em
braided-commutative} with respect to a comodule if it obeys this condition.
Thus a braided group means a Hopf algebra in the braided category equipped with
a class $\CO$ of comodules with respect to which it is braided-commutative.

\begin{propos} Let $B$ be a Hopf algebra in a braided category. Then $B$ coacts
on itself by the braided adjoint coaction defined by
$\Ad=(\id\tens\cdot)(\id\tens
S\tens\id)(\Psi_{B,B}\tens\id)(\Delta\tens\id)\Delta$.
\end{propos}
\proof This is depicted in Figure~1. We have to show that
$(\Ad\tens\id)\Ad=(\id\tens\Delta)\Ad$ (and $\id=(\id\tens\eps)\Ad$, which is
easy and left to the reader). The first diagram on the left in Figure~1 depicts
$(\Ad\tens\id)\Ad$ according to the diagrammatic notation. The upper and lower
parts are each $\Ad$ as stated in the proposition.
Coassociativity of $\Delta$ means that we could combine
$(\Delta\tens\id)\Delta$ as a single vertex with one input and three outputs
(but we should be careful to keep their horizontal order). The first equality
is this coassociativity again and functoriality of $\Psi$ to translate the top
$S$ to the left. The second equality is the fact that $S$ is an anti-coalgebra
homomorphism in the sense $\Delta S=\Psi(S\tens S)\Delta$ (see \cite{Ma:tra}
for a similar fact with regard to the algebra structure). The last equality is
the Hopf algebra axiom that $\Delta$ is an algebra homomorphism to the braided
tensor product algebra, and gives us $(\id\tens\Delta)\Ad$ as required.
\endproof
\begin{figure}
\unitlength=1.05mm
\special{em:linewidth 0.4pt}
\linethickness{0.4pt}
\begin{picture}(116.00,35.00)(-10,0)
\put(13.00,30.00){\makebox(0,0)[cc]{$B$}}
\put(13.00,1.00){\makebox(0,0)[cc]{$B\ \otimes\ B\otimes\ B$}}
\put(45.00,30.00){\makebox(0,0)[cc]{$B$}}
\put(45.00,1.00){\makebox(0,0)[cc]{$B\ \otimes\ B\otimes\ B$}}
\put(81.00,30.00){\makebox(0,0)[cc]{$B$}}
\put(81.00,1.00){\makebox(0,0)[cc]{$B\ \otimes\ B\otimes\ B$}}
\put(116.00,30.00){\makebox(0,0)[cc]{$B$}}
\put(116.00,1.00){\makebox(0,0)[cc]{$B\ \otimes\ B\otimes\ B$}}
\put(28.00,15.00){\makebox(0,0)[cc]{$=$}}
\put(63.00,15.00){\makebox(0,0)[cc]{$=$}}
\put(99.00,15.00){\makebox(0,0)[cc]{$=$}}
\end{picture}
\caption{Proof of braided adjoint coaction}
\end{figure}

For the braided groups $(B,\CO)$ of interest, this canonical braided adjoint
coaction does lie in the class $\CO$, i.e. the Hopf algebra $B$ in the braided
category is braided-commutative with respect to its own braided adjoint
coaction.
One could even require this as an axiom, though we have not done this since the
point of view in \cite{Ma:bg} is more general.

This is a general fact for all braided groups obtained by transmutation of
dual quantum groups $A$, such as of interest in the main text. This process
assigns to a dual quantum group $A$ (with dual quasitriangular structure) a
braided group $B=B(A,A)$ and also to any right $A$-comodule a transmuted right
$B$-comodule in the braided category. Moreover, $B$ is always
braided-commutative with respect to these comodules that arise by
transmutation. They constitute a canonical class $\CO$ in this case. Recall
that transmutation does not change the underlying coalgebra, i.e. the coalgebra
of $B$ coincides
(when the linear spaces are identified) with the coalgebra of $A$, and the
transmutation of comodules is simply to view that same linear map which is an
$A$-comodule, as a $B$-comodule. Noting this, it is not hard to see that the
braided-adjoint coaction in this case is simply the transmutation of the
ordinary quantum adjoint coaction of $A$ on itself.
A similar computation was made for adjoint actions in \cite{Ma:bos}. This is
the fundamental reason that the braided groups that arise by transmutation are
braided-commutative with respect to their own braided adjoint coaction.

\begin{propos} Let $B$ be a Hopf algebra in a braided category and assume that
it is braided-commutative with respect to its own adjoint coaction (e.g. the
braided groups that arise by transmutation). Then $\Ad$ is a comodule algebra
structure in the braided category, i.e.  $\Ad:B\to B\tens B$ is an algebra
homomorphism to the braided tensor product algebra.
\end{propos}
\begin{figure}
\unitlength=1.05mm
\special{em:linewidth 0.4pt}
\linethickness{0.4pt}
\begin{picture}(128.00,90.00)
\put(25.00,3.00){\makebox(0,0)[cc]{$B\ \otimes\ B$}}
\put(25.00,40.00){\makebox(0,0)[cc]{$B\ \otimes\ B$}}
\put(25.00,46.00){\makebox(0,0)[cc]{$B\ \otimes\ B$}}
\put(25.00,85.00){\makebox(0,0)[cc]{$B\ \otimes\ B$}}
\put(59.00,3.00){\makebox(0,0)[cc]{$B\ \otimes\ B$}}
\put(59.00,40.00){\makebox(0,0)[cc]{$B\ \otimes\ B$}}
\put(59.00,46.00){\makebox(0,0)[cc]{$B\ \otimes\ B$}}
\put(59.00,85.00){\makebox(0,0)[cc]{$B\ \otimes\ B$}}
\put(128.00,3.00){\makebox(0,0)[cc]{$B\ \otimes\ B$}}
\put(128.00,40.00){\makebox(0,0)[cc]{$B\ \otimes\ B$}}
\put(128.00,46.00){\makebox(0,0)[cc]{$B\ \otimes\ B$}}
\put(128.00,85.00){\makebox(0,0)[cc]{$B\ \otimes\ B$}}
\put(93.00,3.00){\makebox(0,0)[cc]{$B\ \otimes\ B$}}
\put(93.00,40.00){\makebox(0,0)[cc]{$B\ \otimes\ B$}}
\put(93.00,46.00){\makebox(0,0)[cc]{$B\ \otimes\ B$}}
\put(93.00,85.00){\makebox(0,0)[cc]{$B\ \otimes\ B$}}
\put(42.00,21.00){\makebox(0,0)[cc]{$=$}}
\put(76.00,21.00){\makebox(0,0)[cc]{$=$}}
\put(111.00,21.00){\makebox(0,0)[cc]{$=$}}
\put(42.00,65.00){\makebox(0,0)[cc]{$=$}}
\put(76.00,65.00){\makebox(0,0)[cc]{$=$}}
\put(111.00,65.00){\makebox(0,0)[cc]{$=$}}
\put(9.00,21.00){\makebox(0,0)[cc]{$=$}}
\put(48.00,35.00){\makebox(0,0)[cc]{${\rm Ad}$}}
\put(84.00,35.00){\makebox(0,0)[cc]{${\rm Ad}$}}
\put(120.00,34.00){\makebox(0,0)[cc]{${\rm Ad}$}}
\end{picture}
\caption{Proof that $\Ad$ is an algebra homomorphism}
\end{figure}
\proof The proof is shown in Figure~2. The left-most diagram is $\Ad\circ\cdot$
where $\cdot$ is the multiplication in $B$. The first and second equality use
coassociativity and the Hopf algebra axiom that $\Delta$ is a homomorphism. The
third equality uses the fact that $S$ is an anti-algebra homomorphism (see
\cite{Ma:tra} for a proof). The fourth uses associativity of the multiplication
in $B$ and functoriality to rearrange the diagram so that we can recognise a
part
that is $\Ad$, which we write explicitly in the fifth. The sixth equality  uses
associativity of the multiplication to write in a form suitable for applying
the braided-commutativity condition. Finally, the last equality uses that $B$
is braided-commutative with respect to $\Ad$ to obtain precisely $\Ad\tens\Ad$
followed by the multiplication in $B\tens B$, as required. \endproof

This is the abstract reason for Theorem~3.9. The $B$-comodule algebras in
Propositions~3.7, 3.8 are also obtained by transmutation and hence also lie in
the class with respect to which $B$ there is braided-commutative.

\baselineskip 13pt
\begin{quote}
\noindent Department of Applied Mathematics\\
\noindent\& Theoretical Physics\\
\noindent University of Cambridge\\
\noindent Cambridge CB3 9EW, U.K.
\end{quote}


\end{document}